\documentclass{PoS}
\title{Mini-EUSO experiment to study UV emission of terrestrial and astrophysical origin onboard of the International Space Station}

\ShortTitle{Mini-EUSO on ISS}

\author{\speaker{M. Casolino}$^{e,g,p}$,  M. Battisti$^{a,b}$, A. Belov$^{j}$,   M.~Bertaina$^{a,b}$, F.~Bisconti$^{a}$, S.~Blin-Bondil$^{c}$, F.~Cafagna$^{d}$, G. Cambi\`e$^{e,p}$, F.~Capel$^{f}$, 
 I. Churilo$^{k}$,  G. Cotto$^{a,b}$, A. Djakonow$^{h}$, T.~Ebisuzaki$^{g}$, F. Fausti$^{a,b}$, F.~Fenu$^{a,b}$, C.~Fornaro$^{i}$, A. Franceschi$^{l}$, 
C.~Fuglesang$^{f}$, P.~Gorodetzky$^{m}$, A. Haungs$^{r}$, F.~Kajino$^{n}$, P. Klimov$^{j}$,  L. Marcelli$^{e}$,  W.~Marsza{\l}$^{h}$, M. Mignone$^{a}$, H.~Miyamoto$^{a,b}$,  A. Murashov$^{j}$, T. Napolitano$^{l}$, 
G.~Osteria$^{o}$,  M. Panasyuk$^{j}$, A. Poroshin$^{j}$, E.~Parizot$^{m}$, P. Picozza$^{p,e}$, L.W.~Piotrowski$^{g}$, Z.~Plebaniak$^{h}$, G.~Pr\'ev\^ot$^{m}$, M. Przybylak$^{h}$, E. Reali$^{p}$, {M. Ricci}$^{l}$
 N.~Sakaki$^{g}$, K.~Shinozaki$^{a,b}$, J.~Szabelski$^{h}$, Y.~Takizawa$^{g}$ and M.~Tra\"{i}che$^{q}$, S.~Turriziani$^{g}$, for the JEM-EUSO collaboration.\\
\llap{$^a$}Istituto Nazionale di Fisica Nucleare - Sezione di Torino, Italy, 
\llap{$^b$}Dipartimento di Fisica, Universita' di Torino, Italy,
\llap{$^c$}Omega, Ecole Polytechnique, CNRS/IN2P3, Palaiseau, France,
\llap{$^d$}Istituto Nazionale di Fisica Nucleare - Sezione di Bari, Italy,
\llap{$^e$}Istituto Nazionale di Fisica Nucleare - Sezione di Roma Tor Vergata, Italy,
\llap{$^f$}KTH Royal Institute of Technology, Stockholm, Sweden, 
\llap{$^g$}RIKEN, Wako, Japan,
\llap{$^h$}National Centre for Nuclear Research, Lodz, Poland,
\llap{$^i$}UTIU, Dipartimento di Ingegneria, Rome, Italy,
\llap{$^j$}SINP, Moscow State University, Moscow, Russia,
\llap{$^k$} S.P. Korolev Rocket and Space Corporation «Energia», Korolev, Moscow area, Russia,
\llap{$^l$}Istituto Nazionale di Fisica Nucleare - Laboratori Nazionali di Frascati, Italy,
\llap{$^m$}APC, Univ Paris Diderot, CNRS/IN2P3, CEA/Irfu, Obs de Paris, Sorbonne Paris Cit\'e, France,
\llap{$^n$}Konan University, Kobe, Japan,
\llap{$^o$}Istituto Nazionale di Fisica Nucleare - Sezione di Napoli, Italy,
\llap{$^p$}Universit\'a degli Studi di Roma Tor Vergata - Dipartimento di Fisica, Roma, Italy,
\llap{$^q$}Centre for Development of Advanced Technologies (CDTA), Algiers, Algeria,
\llap{$^r$}Karlsruhe Institute of Technology, Karlsruhe, Germany)\\
E-mail:\email{marco.casolinoi@roma2.infn.it}}

\abstract{Mini-EUSO  will observe  the Earth in the UV range (300 - 400 nm) offering the opportunity to study a variety of atmospheric events such as Transient Luminous
Events (TLEs), meteors and  marine bioluminescence. Furthermore it aims to search for  Ultra High Energy Cosmic Rays (UHECR) above $10^{21}$ eV and Strange Quark Matter (SQM).
The detector is expected to be launched to the International Space Station in August 2019 and  look at the Earth in nadir mode from the UV-transparent window of the Zvezda module of the International Space Station. 
 
The instrument comprises a compact telescope with a large field of view ($44^{\circ}$), based on an optical system employing two Fresnel lenses for light
collection. The light is focused onto an array of 36 multi-anode photomultiplier tubes (MAPMT), for a total of 2304 pixels and the resulting signal is converted 
into digital, processed and stored via
the electronics subsystems on-board. In addition to the main detector, Mini-EUSO contains two ancillary cameras\cite{anccam} for 
complementary measurements in the near infrared (1500 - 1600 nm) and visible (400 - 780 nm) range and also a $8 \times 8$ SiPM imaging array. }

\FullConference{36th International Cosmic Ray Conference -ICRC2019-\\
		July 24th - August 1st, 2019\\
		Madison, WI, U.S.A.}

\begin{document}

\section{Introduction}

\begin{figure}[h!]
 \centering
 \includegraphics[scale=0.3,keepaspectratio=true]{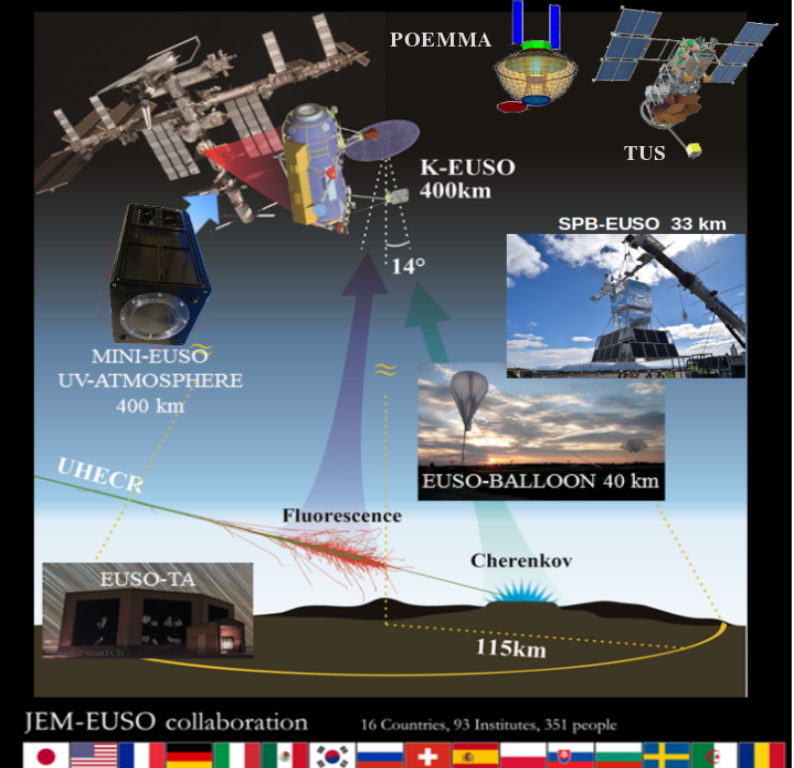}
  \caption{Roadmap of the EUSO telescopes: a) EUSO-TA: Ground detector installed in 2013 at Telescope
Array site, with focal surface installed in 2015. b) EUSO-Balloon: 1st balloon flight from Timmins, CA
(French Space Agency) August 2014; c) NASA Super Pressure Balloon (SPB) flight:2017; d) TUS on
Lomonosov satellite, 2016; e) MINI-EUSO (2019) inside of International Space Station (ISS); f) K-EUSO~\cite{2017ICRC...35..412K}; g) POEMMA~\cite{2019arXiv190706217O}}
 \label{fig:jemeuso_road_map}
\end{figure}

The \textbf{JEM-EUSO} collaboration aims to observe UHECRs from space~\cite{2015ExA....40....3A}.To achieve this goal over 
the years it has realized several detectors (fig.\ref{fig:jemeuso_road_map}): \textbf{EUSO-TA}~\cite{2015ExA....40..301A}, a 
ground-based detector located in front of one of  the Telescope Array fluorescence telescope (2013-); \textbf{EUSO-Balloon}~\cite{Scotti:2016vgy} (2014)
and \textbf{EUSO-SPB1}~\cite{icrc1097} (Super Pressure Balloon) (2017)   two balloon-borne detectors launched respectively from Canada and New Zealand. 
\textbf{EUSO-SPB2}~\cite{spb2_scotti_RG} is in phase of construction for a long duration flight in 2022. These activities are complemented by  \textbf{TUS}~\cite{Klimov:2017lwx}, a Russian 
mission with an array of photomultipliers and 
a mirror optics placed onboard the Lomonosov satellite and launched on April 28th 2016.

\textbf{Mini-EUSO} (UV-Atmosphera in Russian program) is a telescope which will be hosted onboard the ISS ($\sim 400$ km altitude), on a nadir-facing UV transparent 
window, inside the Russian Zvezda module. It is expected to be launched with the Soyuz spacecraft (in an unmanned cargo configuration) from the Baikonur Cosmodrome (Kazakhstan) 
on August 2019. Mini-EUSO is a mission supported by ASI (Italian Space Agency) and ROSCOSMOS (Russian Space Agency) with the JEM-EUSO collaboration contributing to the 
construction of the detector.

\section{Mission Objectives}

\begin{figure}[h!]
\begin{minipage}{16pc}
\includegraphics[width=16pc]{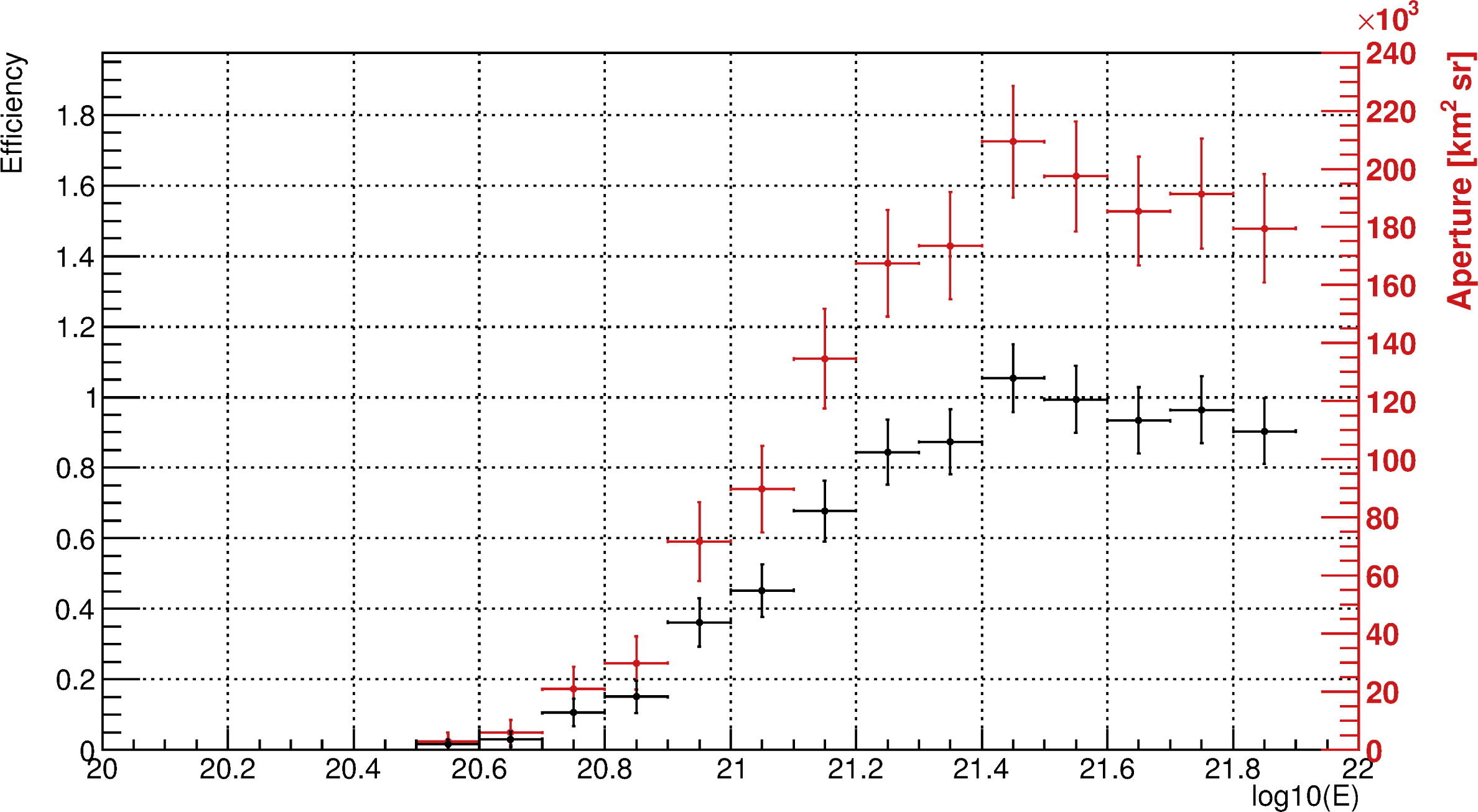}
\caption{\label{eff} Detection efficiency (Left, black) and
geometric aperture (Right, red) as a function of
the EAS energy in eV.~\cite{CAPEL20182954}}
\end{minipage}\hspace{4pc}%
\begin{minipage}{14pc}
\includegraphics[width=12pc]{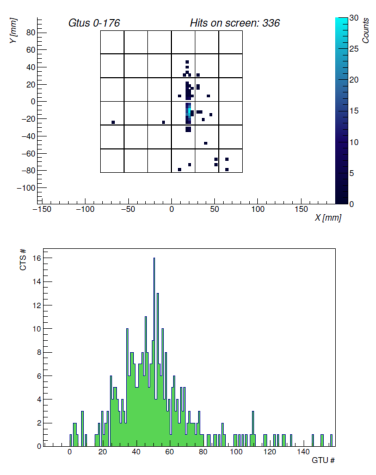}
\caption{\label{eas}Top: UV Photon counts observed in the Mini-EUSO
focal surface for a simulation of an E = $1\times10^{21}$ eV
with an inclination of $80^{\circ}$ to the nadir. Bottom:
Light curve for the same event as a function of time in units of GTU (1 GTU = $2.5 \mu$s).~\cite{2010APh....33..221B}}
\end{minipage}
\end{figure}

The Mini-EUSO~\cite{Casolino:2017fcw} main objective is to observe UHECRs by indirect measurement of the UV fluorescence and Cherenkov light emitted by the $N_2$ molecules in
the Earth atmosphere excited by the interaction with the EAS (Extensive Air Shower) produced by the primary incident particle with an energy greater than $10^{21}$ eV.
 The detection efficiency
and a simulation of an UHECR event signature on the FS made with ESAF (the EUSO Simulation and Analysis Framework~\cite{2010APh....33..221B})
are shown in fig.\ref{eff} and \ref{eas} respectevely.

Regarding the cosmic ray flux, at these energy scale it is crucial to observe the largest possible area.
After TUS, for the first time, Mini-EUSO will perform this measurement from space and this new challenging approach, will provide great improvements to the study of UHECR.
Lens size allows to detect primary cosmic rays with energy above than $10^{21}$ eV threshold.
According to Auger and TA measurements we shouldn't expect to see UHECRs due to the GZK suppression, so we expect
to provide, at least, an upper limit for a null detection with its large exposure. Having a larger exposure than the TUS detector, Mini-EUSO might be able to shed light on the nature of extreme energy EAS-like event recently reported by TUS.  
Moreover, Mini-EUSO will produce a high-resolution map of night-Earth UV emission, focusing on terrestrial emissions~\cite{Casolino:2017fcw}. 
Mini-EUSO will observe many other phenomena: marine phytoplankton bioluminescence, meteoroids with magnitude of $M=+5$ through a online trigger, 
TLEs such as blue jets, sprites and elves events that occur in the upper atmosphere. Also, Mini-EUSO will test space
debris  detection to investigate the possibility of using laser ablation for debris removal~\cite{Ebisuzaki2015102ActaAstronautica}. 
Mini-EUSO will also search for SQM or Strangelets~\cite{2015ExA....40..253A}, a theoretical bound state of equal numbers of up, down, and strange quark. 
According to the strange matter hypothesis, these particles can be produced during UHECR interaction in Earth's atmosphere, but also inside neutron stars or still
having cosmological origin.   

\section{The Telescope}

Mini-EUSO is a compact telescope with dimension $37 \times 37 \times 62 \ cm^3$ and $28$ kg weight. It consists of two main subsystem: Optics and Photo Detector Module (PDM). 
Mini-EUSO optics consists of two PMMA Fresnel lenses ($25$ cm diameter, one of them double sided) which will focus light onto the Focal Surface (FS) with a large 
field of view ($44^{\circ}$).
Mini-EUSO total FS observable area on Earth correspond to $263\times 263 \ km^2$. The  FS consists 
of an array of 36 Hamamatsu $64$ channels Multi-Anode Photomultiplier (MAPMT)  divided into 9 Elementary Cells, for a total of 2304 pixels. The spatial resolution per pixel is 
$0.8^{\circ}$ or $\simeq 5.5\times 5.5 \ km^2$.

\begin{figure}[h!]
\begin{minipage}{14pc}
\includegraphics[width=14pc]{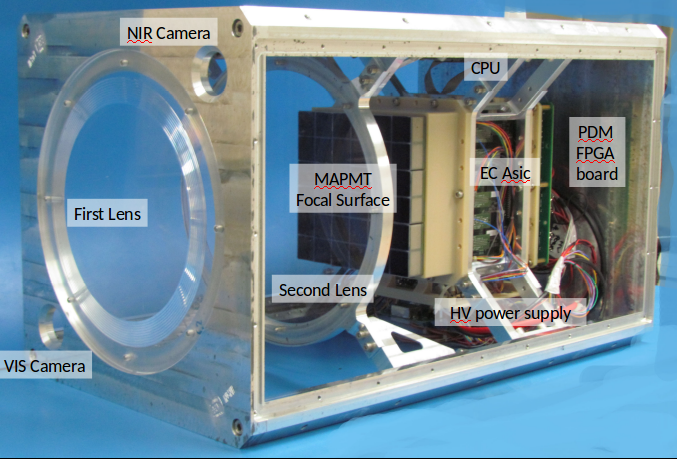}
\caption{\label{instr} Telescope mechanical body with subsystems.}
\end{minipage}\hspace{2pc}%
\begin{minipage}{8pc}
\includegraphics[width=8pc]{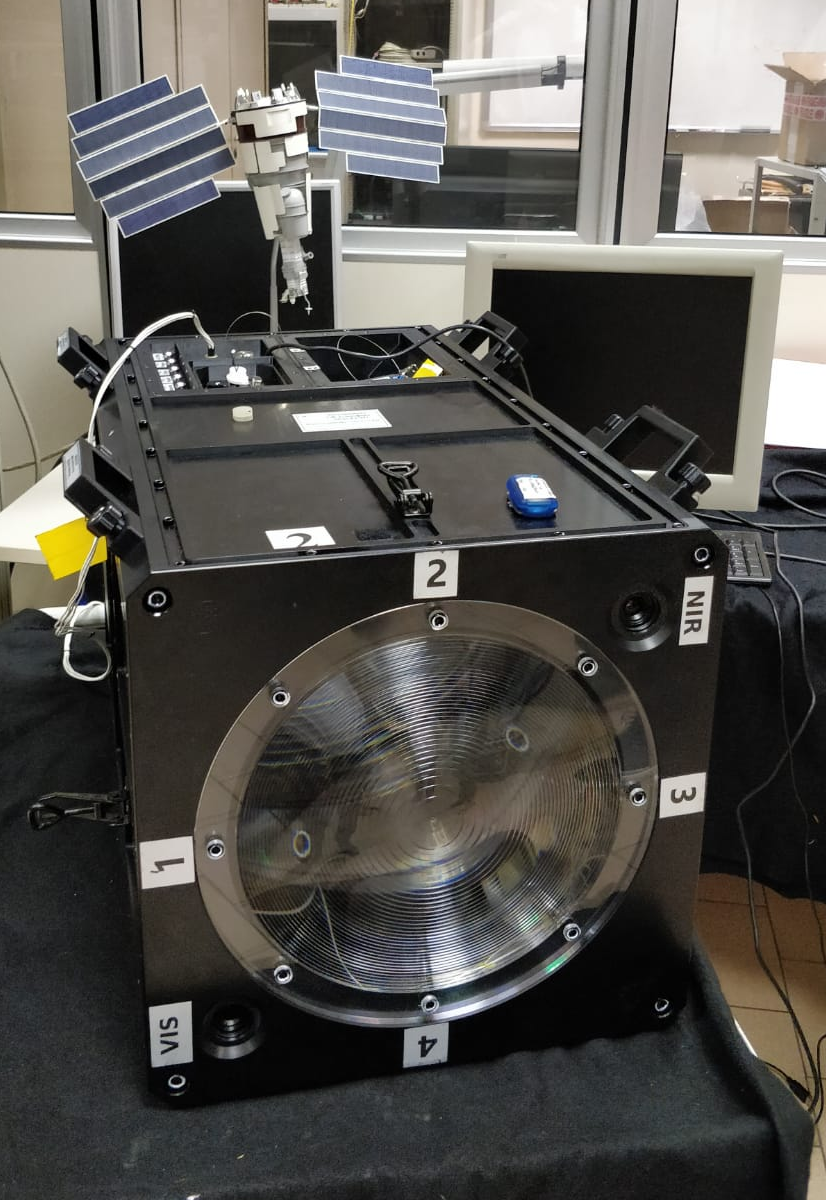}
\caption{\label{FM}Mini-EUSO flight model fully integrated}
\end{minipage}\hspace{2pc}%
\begin{minipage}{9pc}
\includegraphics[width=9pc]{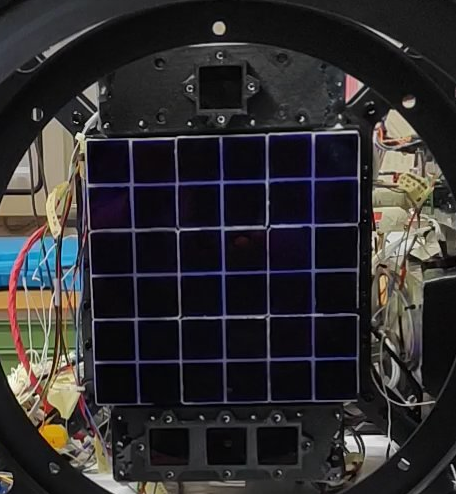}
\caption{\label{focale}Mini-EUSO focal surface.}
\end{minipage} 
\end{figure}

Each MAPMT is powered by a Cockroft-Walton high voltage~\cite{Szabelski_2017} power supply board placed inside a ceramic pad and present a BG3 UV filter on the entry window.
Togheter with FS, the signal and data handling electronics 
form the PDM chain: 6 SPACIROC3~\cite{Blin:2018tjp} (Spatial Photomultiplier Array Counting Integrated ReadOutChip) ASIC boards, 
a Xilinx Zynq XC7Z030 system on chip containing a Kintex7 FPGA with an embedded dual core ARM9 CPU processing system and a PCIe/104 form factor CPU which represent the front end 
electronics. In addition to the main detector, Mini-EUSO 
contains: two ancillary cameras for complementary measurements in the near infrared and visible range, three single pixel UV sensors used as switches for day/night recognition read by
an Atmel 2560 10-bit microcontroller board. Furthermore, Mini-EUSO mounts a 64 channels Multi-Pixel Photon Counter (MPPC) Silicon PhotoMultiplier Tubes (SiPM) C13365 array module 
provided by Hamamatsu Photonics consisting of the photosensitive 
array, a high voltage power supply DC-DC converter, a microcontroller for the bias adjust and temperature-compensation 
tool. A multiplexing board was developed and built on purpose for the MPPC read out while,  is involved on the ancillary sensor read-out. 
The Low Voltage Power Supply (LVPS) boards, consisting of three PCB modules mounting different Vicor DC-DC 
converter, stabilize the $28$ V input voltage coming from ISS providing power for all subsystems and preserving the entire instrumentation from spike and polarization 
inversion. Moreover, the LVPS contains three main bistable relais as housekeeping for the main subsystems connected to a made on purpose board driven by the CPU. 
Mini-EUSO power consumption is around $55$ W and a scheme of the subsystems power distribution is shown below.

\section{Space Qualification Test}

The implementation of the joint Mini-EUSO space experiment requires tests to be carried out on the instrument known as 
General Technical Requirements for Experiment, Equipment and Technical Documents on board ISS. Those can be summarized as follows:
\begin{itemize}
 \item Electro-Magnetic Interference and Conductive (EMC/EMI)
 \item Vibrational and shock
 \item High and Low pressure functionality test
 \item Thermal and humidity 
\end{itemize}

Pressure tests requirements involved the instrument to be placed inside a special chamber at $450 $ mm Hg and $970$ mm Hg for two hours each. Several thermal cylces 
inside a thermal chamber were made, reaching threshold of $\pm 55^ {\circ}$. Requirements were satisfied once by switching on the instrument 
for a functional run after each test. 

\subsection{EMC/EMI Tests}

The aim of the tests is to verify that Mini-EUSO Instrument does not produce any
undesired electromagnetic radiated emissions and that is capable to withstand different
irradiations from external sources. Mini-EUSO engineering model has been subjected to emission and susceptibility tests to demonstrate its electromagnetic
compatibility requirements listed as follows:
\begin{itemize}
 \item Low and High Frequency (LF/HF) Conductive Interference
 \item Electrical Field Intensity Produced by HF Emissions
 \item Pulse Interference
 \item Inrush Current
\end{itemize}

The test equipment is composed of a MXE EMI receiver and spectrum analyzer by Agilent Technologies with a frequency range between $20 \ Hz \div 26.5 \ GHz$, 
a Line Impedance Stabilization Network placed on the power supply for inrush current measurement, ESD Generator,
three different types of antennas: biconical, horn and rod. Mini-EUSO, togheter with antennas were placed inside an anechoic chamber as shown in 
the figures below.

\begin{figure}[h!]
\begin{minipage}{10pc}
\includegraphics[width=10pc]{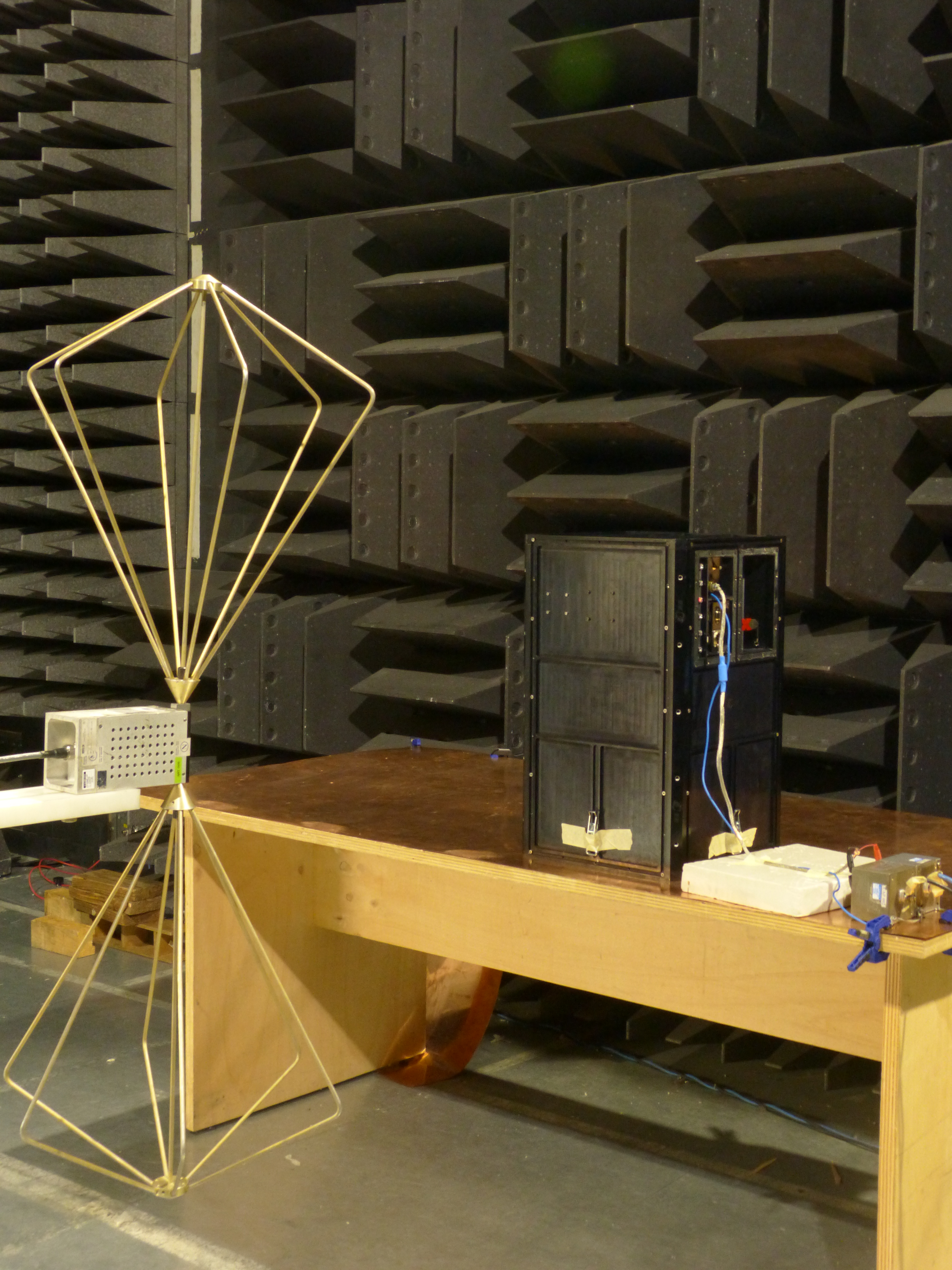}
\caption{\label{emc} Mini-EUSO EMC/EMI qualification tests inside the anechoic chamber}
\end{minipage}\hspace{6pc}%
\begin{minipage}{10pc}
\includegraphics[width=10pc]{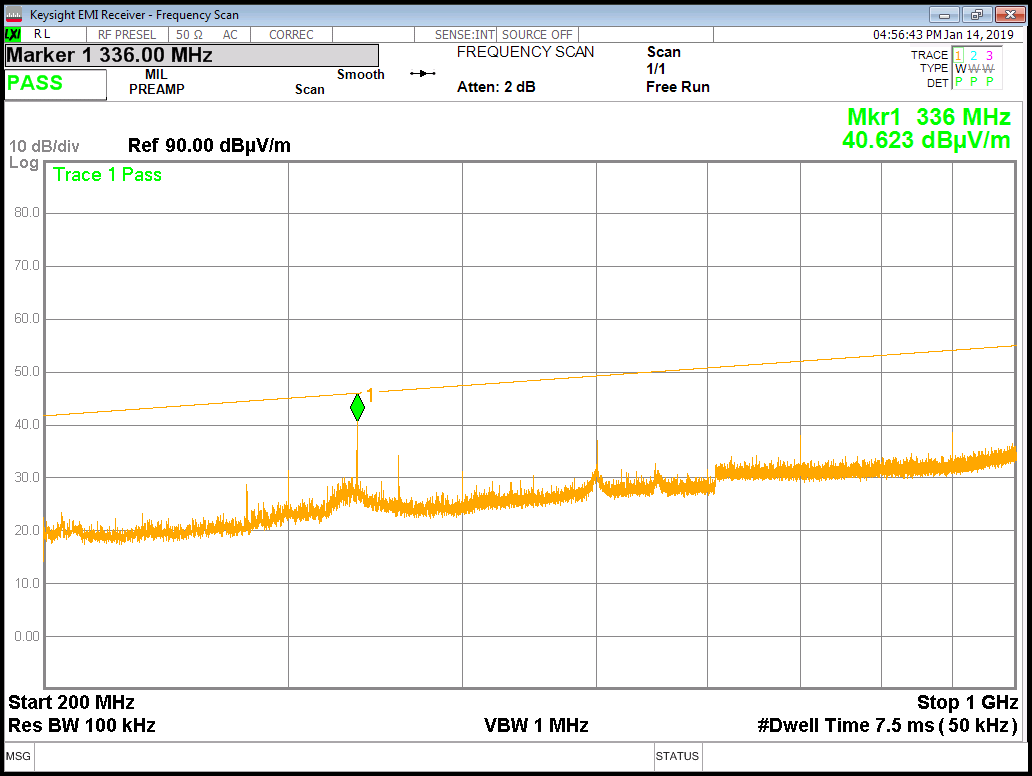}
\caption{\label{hf} Mini-EUSO Conductive Interference spectrum response in range ($200 \div 1000$ MHz)}
\end{minipage}
\end{figure}

During tests antennas have been rotated to obtain the interference spectrum for both vertical and horizontal polarization. All tests have been succesfully
passed. In fig.\ref{hf} are shown the results of the conductive interference in a selected frequencies range. In all frames it's visible a solid
line indicating the conductive interference limits, which is measured in $dB\mu V$.

\subsection{Vibration and Shock Tests}

The Mini-EUSO equipment will be launched into Russian launcher (Soyuz) in hard-
mounted configurations. This qualification procedure required to demonstrate
that Mini-EUSO payloads are able to sustain the random vibration launch loads. In the following will be described the Qualification Vibration Test (QVT) and the 
shock test applied. 
Those tests can be divided into two branches, Random Vibration Tests
and Shock Tests:

\begin{itemize}
 \item Random vibration in the three orthogonal axis X, Y and Z over six different frequencies width
 \begin{itemize}
 \item Resonance survey before all test sessions
  \item Random Insertion (120 seconds) with an overall strenght of $7.42$ g
  \item Random Insertion (480 seconds) with an overall strenght of $3.58$ g
  \item Random Orbital Flight (600 sec) with an overall strenght of $3.84$ g
 \end{itemize}
\item Shock
\begin{itemize}
 \item Seven shock with $3$ ms duration $\pm 40$ g strenght over X, Y and Z axis
 \end{itemize}
\end{itemize}

The criteria for successful testing include a visual inspection after Qualification Vibration and shock Test to show no evidence of
ruptures or damages. Also functional check and resonance comparisons, before and after each test session are required to displacement
of natural frequencies lower than $5 \%$.   

\begin{figure}[h!]
\begin{minipage}{14pc}
\includegraphics[width=14pc]{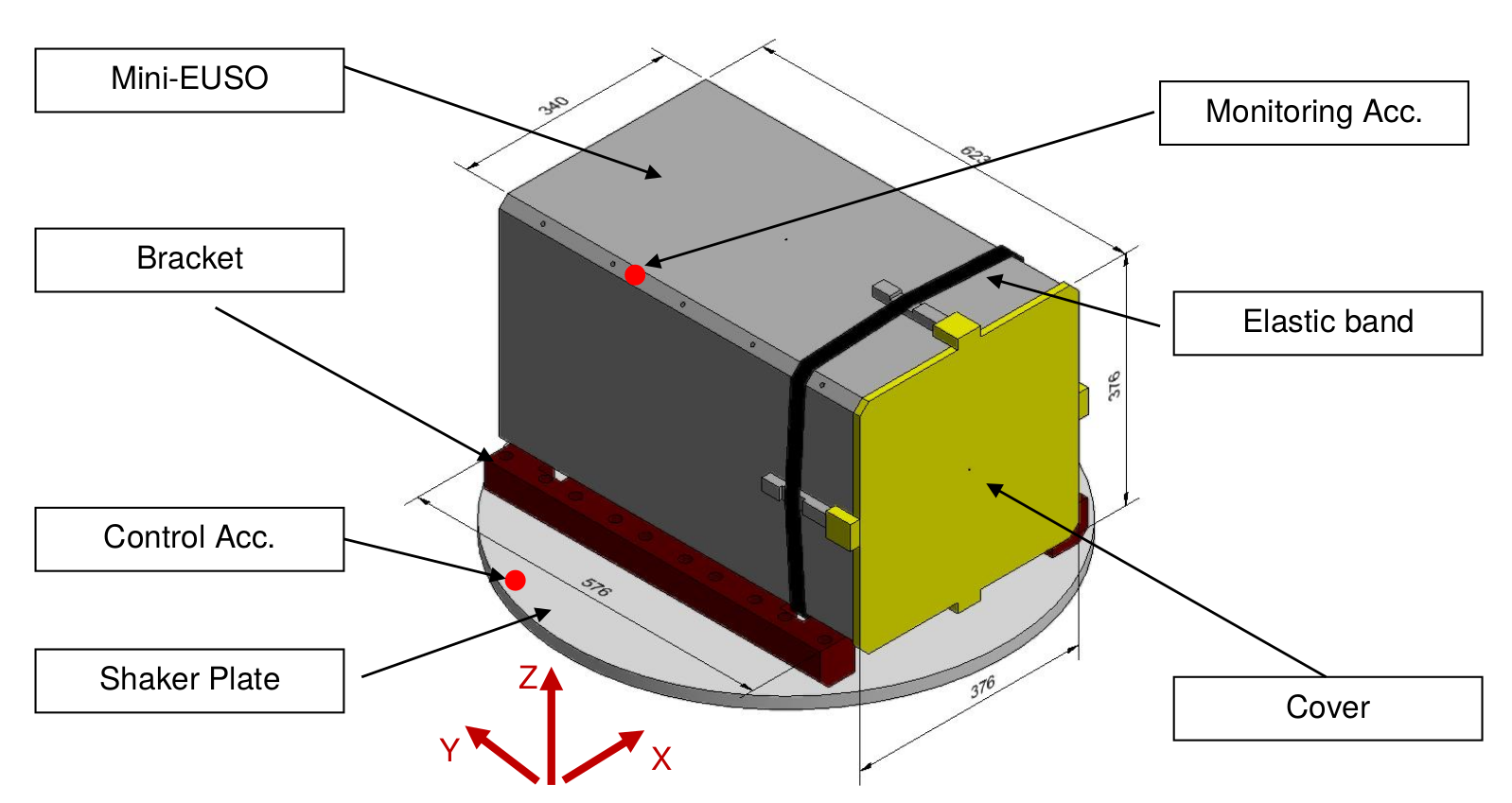}
\caption{\label{vib1} Mini-EUSO Vibration set-up scheme for Z axis}
\end{minipage}\hspace{5pc}%
\begin{minipage}{12pc}
\includegraphics[width=12pc]{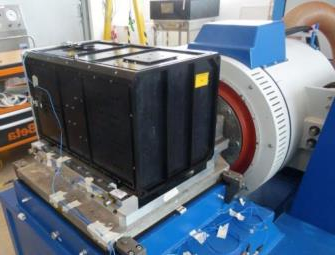}
\caption{\label{vib2}Mini-EUSO accomodation on the shaker - X direction}
\end{minipage}
\end{figure}

 The hard-mounted retaining set-up (shown in fig.\ref{vib1} and \ref{vib2}) is used for the vibration test. 
The hardware under test is fixed to a shaker and on a slip table surface by means a dedicate fixture, two aluminum brackets that
allow to restrain the HW, with M8 screws, to the vibration table. 
All tests have been succesfully passed.

\section{Calibration and Simulations}

\begin{figure}[h!]
\begin{minipage}{14pc}
\includegraphics[width=14pc]{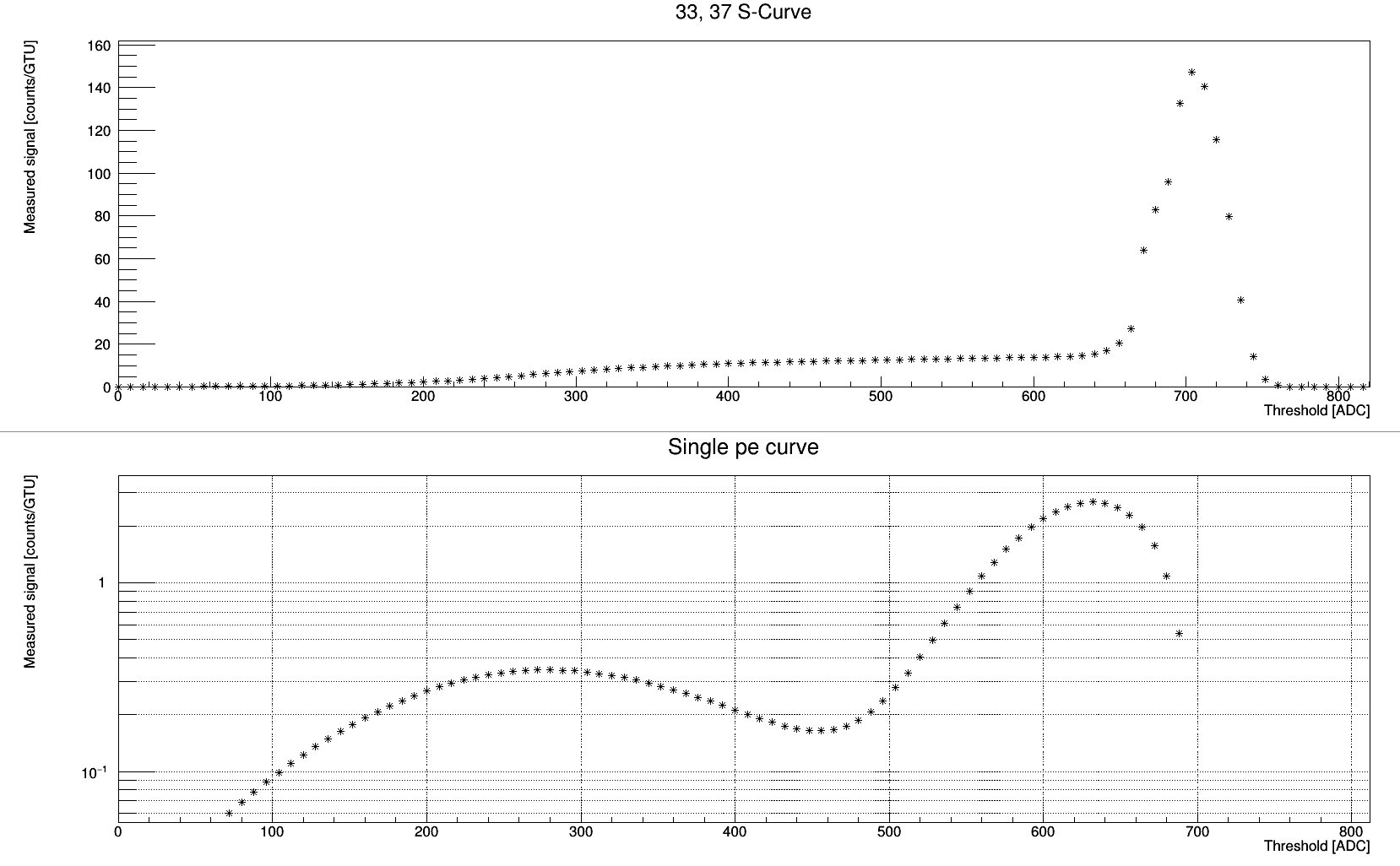}
\caption{\label{scurve}Top:Single pixel S-curve showing counts over a discriminated ADC threshold value. Bottom: S-curve derivative. The threshold for the single photoelectron is placed in the plateau between the left peak and the right pedestal (between 400 and 500 ADC). 
}
\end{minipage}\hspace{2pc}%
\begin{minipage}{14pc}
\includegraphics[width=14pc]{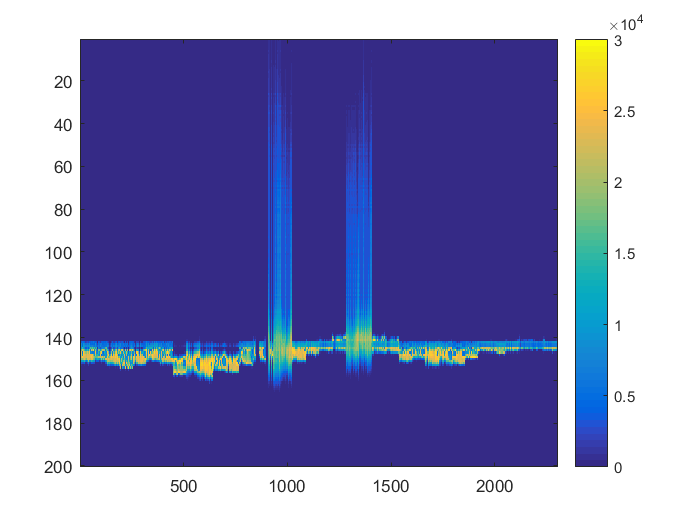}
\caption{\label{allscurve} PDM counts (x-axis) as a function of discrimination threshold (y-axis). The pedestal is at $\simeq  150$ ADC.}
\end{minipage}
\end{figure}

Mini-EUSO is intended to work in a single photoelectron counting mode which 
has advantages over analog measurement in terms of signal-to-noise ratio.
Furthermore, PMT response has to be as uniform as possible. For this reason, the 64 signals from MAPMT anodes are 
fed through the SPACIROC3 preamplifiers which offer adjustable gain to correct the gain non-uniformity of the MAMPT.  
These signals are digitized  and discriminated to count photon triggered pulses and to measure the
photon intensity, thus, allowing to set a threshold for the PMT single photoelectron detection.
If the PMT efficiency and the threshold to separate photon pulses from noise are known, then the number of single 
photoelectron counts is a measurement of the number of single photons incident on the PMT.
The charge or pulse-height spectrum taken at a extremely low light level, such that the response of the 
PMT to a single photon can be measured with high accuracy, is known as a single photoelectron spectrum which is shown for a single pixel in fig.\ref{scurve}.
This show number of discriminated counts per channels corresponding to different pulse height. Each channels is an 
8 bit threshold increment, so that we can distinguish the typical noise pedestal (right) from the rising slope (left) corresponding
to the single photoelectron production needed. 
Visualizing all MAPMT's single photoelectron spectrum response, it is possible also to adjust the gain noting that the pedestals 
are shifted along different channels (fig.\ref{allscurve}). Indeed, the width from the single photoelectron production and the pedestal represent
the PMT's gain, thus all pedestals should occur at the same ADC value to have a uniform PDM. 

{\bf Acknowledgements}
This work was partially supported by Basic Science Interdisciplinary Research Projects of RIKEN
and JSPS KAKENHI Grant (JP17H02905, JP16H02426 and JP16H16737), by the Italian Ministry
of Foreign Affairs and International Cooperation, by the Italian Space Agency through the ASI
INFN agreement n. 2017-8-H.0 and contract n. 2016-1-U.0, by NASA award 11-APRA-0058 in
the USA, by the French space agency CNES, by the Deutsches Zentrum für Luft- und Raumfahrt,
the Helmholtz Alliance for Astroparticle Physics funded by the Initiative and Networking Fund of
the Helmholtz Association (Germany), by
National Science Centre in Poland grant (2015/19/N/ST9/03708 and 2017/27/B/ST9/02162),  and by State Space Corporation ROSCOSMOS and
Russian Foundation for Basic Research (grant 16-29-13065).


\begin{thebibliography}{10}

\bibitem{2015ExA....40....3A}
J.~H. {Adams},   et~al.
\newblock {The JEM-EUSO mission: An introduction}.
\newblock {\em Experimental Astronomy}, 40:3--17, November 2015.

\bibitem{2015ExA....40..301A}
J.~H. {Adams}, 
  et~al.
\newblock {Ground-based tests of JEM-EUSO components at the Telescope Array
  site, ''EUSO-TA''}.
\newblock {\em Experimental Astronomy}, 40:301--314, November 2015.

\bibitem{Scotti:2016vgy}
V.~Scotti and G.~Osteria.
\newblock {EUSO-Balloon: The first flight}.
\newblock {\em Nucl. Instrum. Meth.}, A824:655--657, 2016.
\bibitem{anccam}
S. Turriziani et al., 
\newblock {Secondary cameras onboard the Mini-EUSO experiment: Control software and calibration}.
\newblock {\em ASR}, 64, 5, 1188-1198, 2019.
\bibitem{icrc1097}
L.~Wiencke.
\newblock Euso-spb1 mission and science.
\newblock {\em PoS}, ICRC2017:1097, 2017.

\bibitem{spb2_scotti_RG}
V.~Scotti and G.~Osteria.
\newblock The euso-spb2 mission.
\newblock {\em Nuclear Instruments and Methods in Physics Research Section A:
  Accelerators, Spectrometers, Detectors and Associated Equipment}, 05 2019.

\bibitem{Klimov:2017lwx}
P.~A. Klimov et~al.
\newblock {The TUS detector of extreme energy cosmic rays on board the
  Lomonosov satellite}.
\newblock {\em Space Sci. Rev.}, 212(3-4):1687--1703, 2017.

\bibitem{2017ICRC...35..412K}
P.~{Klimov}, M.~{Casolino}, and {JEM-EUSO Collaboration}.
\newblock {Status of the KLYPVE-EUSO detector for EECR study on board the ISS}.
\newblock {\em International Cosmic Ray Conference}, 301:412, Jan 2017.

\bibitem{2019arXiv190706217O}
A.~V. {Olinto}, et al,
\newblock {POEMMA (Probe of Extreme Multi-Messenger Astrophysics) design}.
\newblock {\em arXiv e-prints}, page arXiv:1907.06217, Jul 2019.

\bibitem{Casolino:2017fcw}
M.~Casolino, et a.,
\newblock {Science of Mini-EUSO detector on board the International Space
  Station}.
\newblock {\em PoS}, ICRC2017:369, 2018.

\bibitem{2010APh....33..221B}
C.~{Berat}, et al,
\newblock {Full simulation of space-based extensive air showers detectors with
  ESAF}.
\newblock {\em Astroparticle Physics}, 33(4):221--247, May 2010.

\bibitem{CAPEL20182954}
F.~Capel, et al.,
\newblock Mini-euso: A high resolution detector for the study of terrestrial
  and cosmic uv emission from the international space station.
\newblock {\em Advances in Space Research}, 62(10):2954 -- 2965, 2018.
\newblock Origins of Cosmic Rays.

\bibitem{Ebisuzaki2015102ActaAstronautica}
T.~Ebisuzaki, M.~N. Quinn, S.~Wada, l.~W. Piotrowski, Y.~Takizawa, M.~Casolino,
  M.~Bertaina, P.~Gorodetzky, E.~Parizot, T.~Tajima, R.~Soulard, and G.~Mourou.
\newblock Demonstration designs for the remediation of space debris from the
  international space station.
\newblock {\em Acta Astronautica}, 112:102 -- 113, 2015.

\bibitem{2015ExA....40..253A}
J.~H. {Adams}, 
  et~al.
\newblock {JEM-EUSO: Meteor and nuclearite observations}.
\newblock {\em Experimental Astronomy}, 40:253--279, November 2015.

\bibitem{Szabelski_2017}
Z.~Plebaniak, et al.,
\newblock {HVPS system for EUSO detectors}.
\newblock {\em PoS}, 301:378, 2017.

\bibitem{Blin:2018tjp}
S.~Blin, et al.,
\newblock {SPACIROC3: 100 MHz photon counting ASIC for EUSO-SPB}.
\newblock {\em Nucl. Instrum. Meth.}, A912:363--367, 2018.

\bibitem{refId0}
{Cambi\`e, G.} and {Marcelli, L.}
\newblock Integration and testing of the mini-euso telescope.
\newblock {\em EPJ Web Conf.}, 209:01047, 2019.

\end{thebibliography}
\end{document}